%% file: main.tex
\documentclass[conference]{IEEEtran}
\IEEEoverridecommandlockouts
\usepackage{cite}
\usepackage{comment}
\usepackage{amsmath,amssymb,amsfonts}
\usepackage{textcomp}
\usepackage[table]{xcolor}
\usepackage{arydshln}
\usepackage{tablefootnote}
\usepackage[export]{adjustbox}
\usepackage{wrapfig}
\usepackage{graphicx}
\usepackage{algpseudocode} 
\usepackage[mathscr]{eucal}
\usepackage{amsmath}
\usepackage{multicol}
\usepackage{algorithm} 
\usepackage{bm}
\usepackage{hyperref}
\usepackage{booktabs}
\usepackage[applemac]{inputenc}
\usepackage{xparse}
\usepackage{xifthen}
\usepackage{mathtools}
\usepackage{multirow}
\usepackage{textcomp}
\usepackage{todonotes}
\usepackage{pifont}
\usepackage{soul}
\usepackage{color}
\usepackage{upgreek}
\usepackage{footnote}
\makesavenoteenv{tabular}
\usepackage{subcaption}
\usepackage{lipsum}
\usepackage{xcolor}
\definecolor{anti-flashwhite}{rgb}{0.95, 0.95, 0.96}

\usepackage[cp858]{inputenx}
\renewcommand{\textsl}{\textit}
\newcommand{\comm}[1]{}

\NewDocumentCommand{\vect}{ O{} O{} m }{\mathbf{#3}\ifthenelse{\isempty{#1}}{}{^{(#1)}}\ifthenelse{\isempty{#2}}{}{_{#2}}}

\NewDocumentCommand{\mat}{ O{} O{} m }{\mathbf{#3}\ifthenelse{\isempty{#1}}{}{^{(#1)}}\ifthenelse{\isempty{#2}}{}{_{#2}}}

\NewDocumentCommand{\ten}{ O{} O{} m }{\pmb{\mathscr{#3}}\ifthenelse{\isempty{#1}}{}{^{(#1)}}\ifthenelse{\isempty{#2}}{}{_{#2}}}

\makeatletter
\newcommand{\thickhline}{%
    \noalign {\ifnum 0=`}\fi \hrule height 1pt
    \futurelet \reserved@a \@xhline
}

\begin{document}


\title{Interactive Distillation of Large Single-Topic Corpora of Scientific Papers}

 \author{
 \IEEEauthorblockN{Nick Solovyev}
 \IEEEauthorblockA{\textit{Theoretical Division, LANL} \\
 Los Alamos, USA \\
 nks@lanl.gov}
 \\
 \IEEEauthorblockN{Maksim E. Eren}
 \IEEEauthorblockA{\textit{Advanced Research in Cyber Systems, LANL} \\
 Los Alamos, USA \\
 maksim@lanl.gov}
 \and
 \IEEEauthorblockN{Ryan Barron}
 \IEEEauthorblockA{\textit{Theoretical Division, LANL} \\
 Los Alamos, USA \\
 barron@lanl.gov}
 \\
 \IEEEauthorblockN{Kim \O. Rasmussen}
 \IEEEauthorblockA{\textit{Theoretical Division, LANL} \\
 Los Alamos, USA \\
 kor@lanl.gov}
 \and
 \IEEEauthorblockN{Manish Bhattarai}
 \IEEEauthorblockA{\textit{Theoretical Division, LANL} \\
 Los Alamos, USA \\
 ceodspspectrum@lanl.gov}
 \\
 \IEEEauthorblockN{Boian S. Alexandrov}
 \IEEEauthorblockA{\textit{Theoretical Division, LANL} \\
 Los Alamos, USA \\
 boian@lanl.gov}
}
\maketitle

\begin{abstract}
Highly specific datasets of scientific literature are important for both research and education. However, it is difficult to build such datasets at scale. A common approach is to build these datasets reductively by applying topic modeling on an established corpus and selecting specific topics. A more robust but time-consuming approach is to build the dataset constructively in which a subject matter expert (SME) handpicks documents. This method does not scale and is prone to error as the dataset grows. Here we showcase a new tool, based on machine learning, for constructively generating targeted datasets of scientific literature. Given a small initial "core" corpus of papers, we build a citation network of documents. At each step of the citation network, we generate text embeddings using the transformer generated science-specific large language model SciNCL [Ostendorff, Malte, et al. "Neighborhood contrastive learning for scientific document representations with citation embeddings." arXiv preprint arXiv:2202.06671 (2022).] 
and visualize the embeddings through dimensionality reduction. Papers are kept in the dataset if they are "similar" to the core or are otherwise novelly pruned through human-in-the-loop selection. Additional insight into the papers is gained through sub-topic modeling using SeNMFk. We demonstrate our new tool for literature review by applying it to two different fields in machine learning.
\end{abstract}


\section{Introduction}
\input{sec_introduction}

\section{Related Work}
\label{sec:relevant_work}
\input{sec_relevant_work}

\section{Method}
\label{sec:method}
\input{sec_method}

\section{Results}
\label{sec:results}

\input{sec_results}

\section{Conclusion}
\label{sec:conclusion}
\input{sec_conclusion}

\section{Acknowledgment}
This research was funded by DOE National Nuclear Security Administration (NNSA) - Office of Defense Nuclear Nonproliferation R\&D NA-22 grant DE-AC52-06NA2539 supported by Los Alamos National Laboratory's Institutional Computing Program, and by the U.S. Department of Energy National Nuclear Security Administration under Contract No. 89233218CNA000001.

\bibliographystyle{IEEEtran}
\bibliography{main}

\end{document}

%% file: sec_introduction.tex
One of the integral tasks of scientific research is the literature review of highly specific topics of interest. Literature review often involves identifying papers of interest based on keyword searches and following the relevant citations. This manual process, however, is prone to miss potentially significant papers and information. In addition, organizing highly specific scientific literature datasets and applying data analysis techniques, such as topic modeling, may allow a deeper understanding of a given field and the discovery of new research directions. However, curating such highly-specific datasets of scientific literature requires the time-consuming help of a subject matter expert (SME). Here, we introduce a new assistant tool based on machine learning (ML) that allows for building highly-specific scientific literature datasets. Bibliographic Utility Network Information Expansion (BUNIE) streamlines the literature review task with a user-friendly and intuitive system while enhancing the specificity of the papers of interest using ML techniques and integrated human-in-the-loop procedures.

In this work, we contribute a novel approach to the scientic dataset expansion problem by jointly integrating Transformer-based document text embeddings with human-in-the-loop pruning to generate targeted scientific datasets. We then use non-negative matrix factorization (NMF) with automatic model determination (NMFk) for modeling the topics in these papers to further refine our datasets \cite{SmartTensors}. Our approach is unique in its inclusion of a human-in-the-loop for enhancing and distilling the extracted topics, such that the corpus of papers is narrowed down via an interactive process. To the best of our knowledge, this iterative method is the first of its kind to offer users the ability to analyze the topic modeling results and apply their feedback to enhance the literature review procedure by steering the ML output. The feedback loop enables the users to grow and refine the results until a targeted dataset of a specific size is reached, providing a unique and interactive solution to large-scale literature review.


The process begins with a small number of core papers selected from a topic of interest by an SME. At this initial stage, the topic may not fully align with the user's specific objectives and is likely incomplete. The core papers are used as a reference to obtain an additional set of relevant documents that increase the size and enhance the specificity of the existing dataset. The additional documents are selected using a citation network formed from the existing papers in the dataset. The expansion results are then pruned using multiple methods, including an interactive selection by the user, document embedding similarity metrics, and topic modeling.
In contrast to the traditional static approach of computing the topics, our approach is iterative and dynamic. It allows repetition of this refinement cycle, growing the dataset with each iteration. This enables the creation of large but specific datasets, ideal for training large language models. Through this interactive, user-driven approach, we empower users to steer the topic extraction process directly, ensuring the results are tailored to their specific requirements. This paper demonstrates our novel tool by exploring the scientific literature on applying tensor decomposition for numerical solutions of partial differential equations.

Our contributions include:
\begin{itemize}
    \item Introducing a novel paper selection and visualization tool for scientific dataset curation and literature review.
    \item Utilizing text embeddings together with dimensionality reduction techniques to model the documents.
    \item Integrating our machine learning approach to scientific literature with human-in-the-loop procedures for refining and guiding text modeling.
    \item Demonstrating the capabilities of our tool by applying it to the scientific literature in two different scientific fields.
\end{itemize}

\begin{figure*}[ht!]
    \centering
    \includegraphics[width=.9\textwidth]{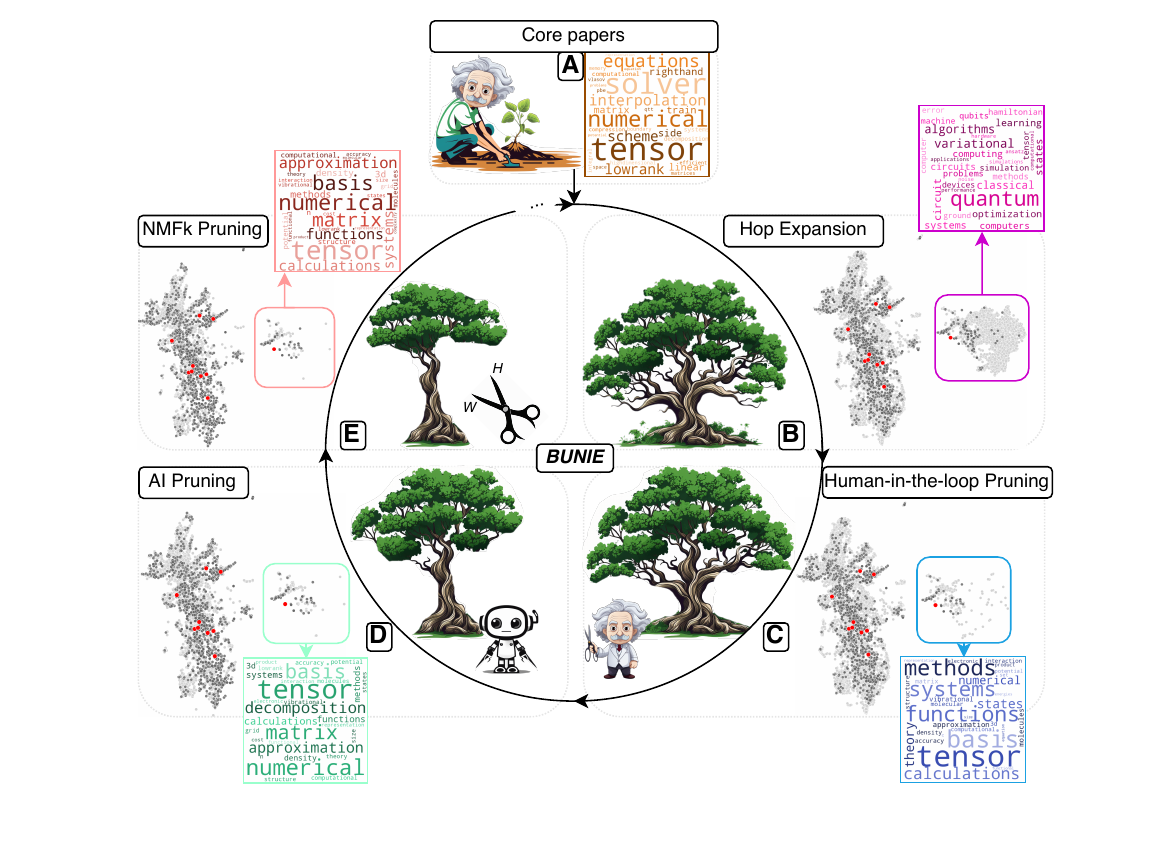}
    \caption{Illustration of BUNIE's distillation pipeline.
    \textbf{Panel A} inputs to BUNIE the SME-selected, highly-specific papers (core), where a subset is represented as a bag-of-words wordcloud.  
    \textbf{Panel B} dataset is expanded through the core's citation network. A subset of papers is selected to showcase how much the content differs from the core
    \textbf{Panel C} features human-in-the-loop pruning. Reduction in the subset cluster is visible and the wordcloud begins to resemble the core. 
    \textbf{Panel D} documents are pruned through a document embedding heuristic, removing papers too far from the core in the embedding space. The dataset becomes more compact and the subset begins to approach wordcloud parity with the original papers. 
    \textbf{Panel E} topic modeling through SeNMFk further prunes the dataset. H-clustering the factorization removes clusters that lack core papers, producing to a neatly trimmed tree. This final stage has a dense set yet numerous documents indicating a successful distillation, evidenced in a wordcloud closely resembling the original. Additional cycles can be made for refinement or the data extracted in repose for downstream analysis, the option as the \textbf{ellipse}.}
    \label{fig:overview}
    
\end{figure*}
    

%% file: sec_relevant_work.tex
This section summarizes techniques and prior works applied to forging a highly-specific dataset of research papers.



\emph{1) Topic Modeling \& Tensor Decomposition:} A  common approach to topic modeling is through Non-Negative Matrix Factorization (NMF) \cite{Paatero1994PositiveMF, doc_cluster_nmf}. NMF can be applied to a words by documents matrix to identify latent patterns within the corpus. An extension to NMF, Semantic NMF with automatic model determination (SeNMFk), is leveraged in \cite{vangara2020semantic,vangara2021finding,seNMFK_lanl} to perform topic modeling while incorporating the text's semantic structure. The aformentioned documents by word matrix used in NMF and SeNMFk is usually the term frequency\textendash inverse document frequency (TF\textendash IDF) matrix together with the co-occurrence/word-context matrix, the values of which represent the number of times two words co-occur in a predetermined window of the text. This is a common method of vectorization, however, more advanced methods to process the documents exist.

\emph{2) Document Embeddings \& Transformers:}
Vector representations of a  text were previously used for dimensional mapping, cross-comparisons, and similarity analysis \cite{pmlr-v37-kusnerb15}. 
Common models for learning word embeddings have been Global Vectors for Word Representation (GloVe)\cite{pennington2014glove} and Word2Vec \cite{mikolov2013efficient}. Recently, transformers have been used for large language models (LLM) as internal states, as well as for topic modeling \cite{9053399, lo2021transformer, zhu2021topicdriven, bert-topic_approaches}. A popular tranformer-based LLM is the Bidirectional Encoder Representations from Transformers (BERT) \cite{DBLP:journals/corr/abs-1810-04805}. An example of BERTs topic modeling and sentiment analysis is discussed in Ref. \cite{Praveen2023UnderstandingTP}, where emotions related to the use of ChatGPT \cite{openai2021} are studied through social media posts by medical field researchers. While transformers, as used in BERT are useful, they have limitations, as demonstrated in Ref. \cite{9003958}, which addresses BERT's few hundred-word input capacity by adding transformer layers to segment text, paired with two activation layers for final classification. 
In our work, we apply the SciNCL transformer to generate text embeddings \cite{ostendorff2022scincl}. Citation data is used as an additional training signal in SciNCL's document embedding closeness, aiding the document distance determinations.  

\emph{3) Data Visualization and Tools:} 
A range of tools is available to explore and analyze research papers. For instance, citation network and topic modeling tools such as Topic Modeling Tool \cite{topic_model_tool_1}, 
and Stanford Topic Modeling Toolbox \cite{topic_model_tool_2} are publicly available. While these tools excel in gathering topical data from their inputs, they lack visual representation.
Another tool with visualizations, 'Connected Papers,' serves as a resource for discovering scientific literature~\cite{connectedPapers_gui}. From single document inputs, Connected Papers produces a graph where each paper is a node, positioned according to a coupling of the co-citations and bibliography rather than direct citations \cite{connectedPapers_gui}. While 'Connected Papers' is useful for exploring scientific literature based on the bibliography information, our tool advances the utility of bibliography information by creating a specialized dataset of documents leveraging a citation network coupled with human-in-the-loop and machine learning procedures.
Another research paper visualization tool, designed specifically for the influx of Covid-19 papers at the height of the pandemic, is explained in Ref. \cite{10.1145/3395027.3419591}. Within this tool, the process begins with text cleaning (tokenization, removal of stop-words, \& punctuation \& capitalization),  transformation into a TF-IDF matrix, then  t-distributed stochastic neighbor embedding (t-SNE) \cite{van2008visualizing} reduced dimensions for graphing. Here, we use Uniform Manifold Approximation and Projection (UMAP) \cite{mcinnes2020umap} to reduce the 768-dimensional embeddings output by SciNCL \cite{ostendorff2022scincl} to a two-dimensional projection. 

\emph{4) Human in the Loop:}
Incorporating user feedback into the systems has seen recent adoption into several schemes, including OpenAI's ChatGPT \cite{openai2021} and Google's BARD \cite{googleBard}. 
In the study Ref. \cite{Human_in_the_loop_Graph}, a knowledge graph is built by the framework textually prompting a user, collecting feedback in every response to provide an acceptable retail-item recommendation. 
Significant differences between BUNIE and Ref. \cite{Human_in_the_loop_Graph} exist. For instance, the study's structure provides one recommendation, whereas BUNIE offers an entire dataset. Interactive modes also differ. Our system uses click-and-drag selection to delete papers rather than a textual conversation. 
Furthermore, BUNIE only removes papers at the HITL phase until further citation hops are requested for more documents. Contrastly, the tool described in Ref. \cite{Human_in_the_loop_Graph} requests positive and negative feedback about recommendations. 
\begin{figure*}[ht!]
    \centering
    \includegraphics[width=.8\textwidth]{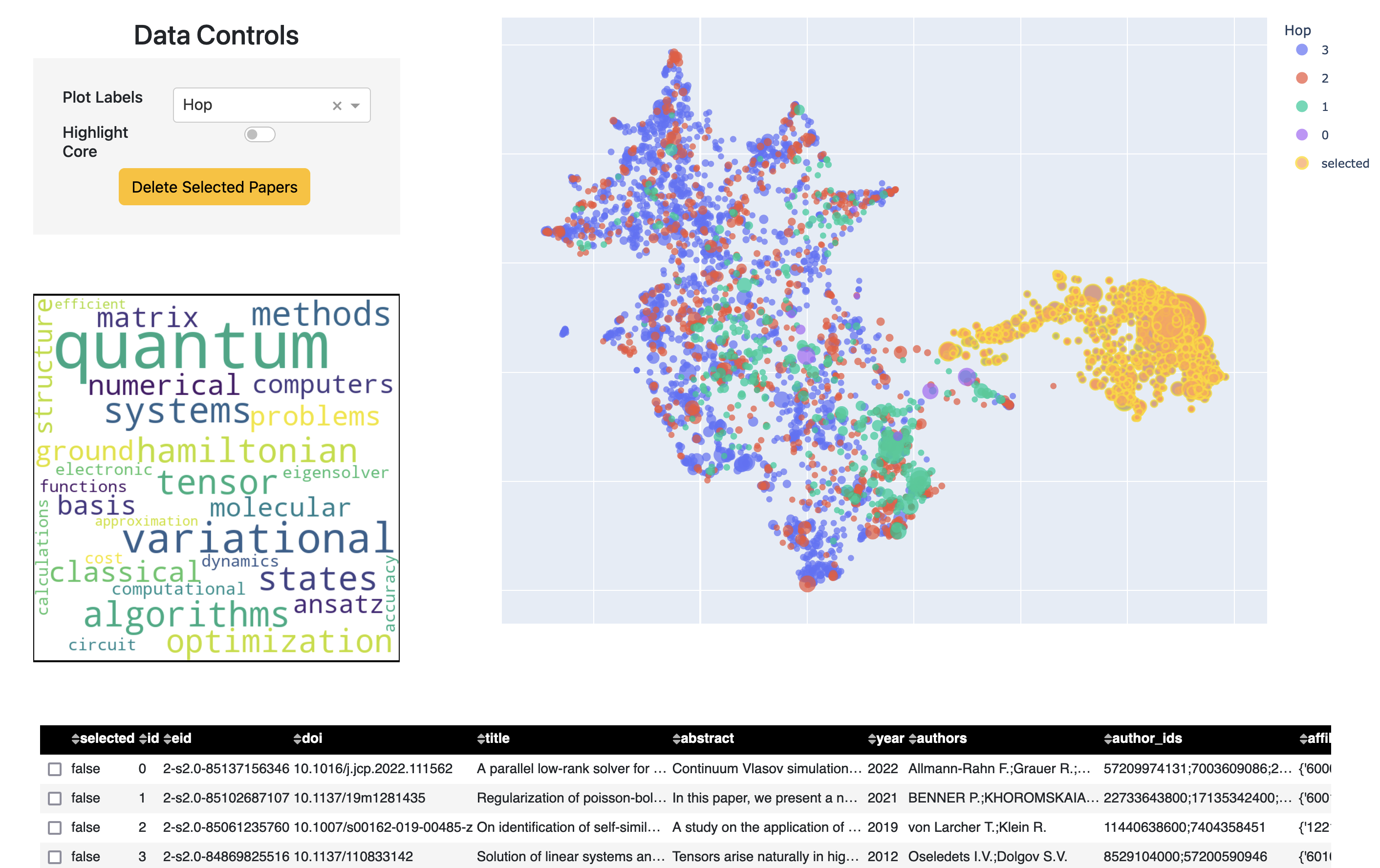}
    \caption{Screenshot of the GUI built for BUNIE. The user is able to upload core papers, perform hops, and do the HITL pruning. This screenshot shows a user pruning a 3-hop dataset. The papers on the right-hand side of the plot (highlighted in yellow) have been manually selected, and the bag-of-words wordcloud for the selection is shown to the left. The \textit{Hop} legend represents the number of hops (hop 0 is the core set of documents), and the SME selected/highlighted documents.}
    \label{fig:gui_example}
\end{figure*}
A HITL work more similar to BUNIE is described in Ref. \cite{image_dataset_human_in_loop}, which aims to build labeled image datasets for Computer Vision (CV) applications.  A user labels a few images, which extrapolate to all in the image's cluster, then the images are evaluated by a model for reassignment. Like BUNIE, the process is iterated to convergence but differs in direct user influence of datum retention. 
Moreover, HITL has been examined from the perspective of assisting artificial intelligence (AI) in Ref. \cite{human_in_loop_survey}, specifically for natural language processing (NLP), CV, an NLP and CV pairing, and real-world robotic applications. Still, HITL from the perspective of non-robotic real-world applications supported by NLP is not considered. In our tool, a user inputs one or more documents, iteratively grows the data, and deletes documents at every iteration, hand-in-hand with both AI and  tensor pruning.

%% file: sec_method.tex

The utility of BUNIE comes from the combination of being able to quickly expand a dataset of publications by traversing the citation network in combination with being able to curate the dataset at scale by effectively using document text embeddings. As depicted in Figure~\ref{fig:overview}, the workflow is cyclical as it involves iterative steps of acquiring new papers through the citation network and then refining this expanded dataset using various pruning techniques. The ultimate goal is to create an extensive collection of scientific literature centered around a specific topic, using a small, hand-picked set of relevant papers as the starting point. 
\subsubsection{\textbf{Selecting the Core}} 
 First, the user provides BUNIE with a set of "core" papers, comprised of a unifying theme or topic, as the foundation of the dataset.
A subject matter expert (SME) should select and/or review the core for the best results to ensure quality and relevance. It is important to remember that BUNIE expands the dataset by traversing the citation network of the core papers. 
A single, well-cited document may produce an extensive dataset after a few iterations/hops following the citation network, while a collection of less frequently cited documents might yield a more limited network. 
 In our experiments, BUNIE has been successfully applied to cores ranging from as few as 6 documents and as many as 63 documents. The core papers are inputted  into BUNIE using unique paper identifiers such as DOI. Using the SemanticScholar API \cite{Kinney2023TheSS}, BUNIE extracts basic information about these documents, including the title, abstract, year of publication, authors, citations, and references.
\subsubsection{\textbf{Expanding the Dataset}} 
With the core established, the user can grow the dataset by making a "hop" within the citation network. The citation network represents a directed graph formed by publications and their respective citations. If we denote a document, $a$, as belonging to a set of documents $X$ and a document, $b$, belonging to the set of their citation $X^c$, we can say that $a \rightarrow b$ if and only if $b$ cites $a$. In this context, a hop can be defined as \( X := X \cup X^c \). In this fashion, a second hop would also incorporate the citations from the documents in $X^c$, which was acquired from the first hop. The number of hops performed is left to the user's discretion, thereby controlling the scale of dataset expansion. The process can continue until the dataset reaches a desired size or until the entire citation network has been traversed. BUNIE also offers the capability to form the citation network with the edges reversed, using references as the basis instead of citations. This feature can be particularly useful when the core consists of relatively new or infrequently cited publications. 

\subsubsection{\textbf{Pruning the Dataset}} Given the interconnected nature of the citation network, not every paper found through the hop process will be relevant to the core. For example, a highly influential publication may be cited as an acknowledgment in subsequent studies focusing on entirely new issues. Thus, it is crucial to perform pruning at each hop along the citation network to prevent irrelevant topics from propagating within the growing dataset. In BUNIE, pruning is accomplished through a combination of the following three techniques.  

\subsection{Human in the Loop (HITL) Pruning: } Textual similarity comparison presents a substantial challenge for humans and computational algorithms.  To simplify this task, we employ SciNCL \cite{ostendorff2022scincl} to transform the aggregated titles and abstracts of the dataset into 768-dimensional embeddings. These high-dimensional embeddings are reduced to a two-dimensional projection using UMAP \cite{mcinnes2020umap}. 
Semantically similar papers tend to cluster together in this two-dimensional space when plotted on a scatter plot, providing an intuitive visual representation of the dataset's structure. Although not perfect, this process simplifies manual content comparison. 

To aid in the manual analysis and document pruning, we have designed a graphical user interface (GUI) to quickly select and examine many papers using the UMAP visualized projection of the embeddings, as shown in Figure~\ref{fig:gui_example}. The SME can highlight papers by drawing a custom lasso or rectangle over the projected papers. The tool then generates a bag-of-words wordcloud to show the most frequent vocabulary in the chosen paper set. For a finer-grain analysis, the GUI provides a data table displaying all known data fields for the selected papers that an SME can analyze. 

\subsection{Automatic Pruning of Document Embeddings} As with any dimensionality reduction algorithm, UMAP necessitates tradeoffs in the data's representation in the two-dimensional space. While useful for document visualization and enabling HITL pruning, a significant portion of the embedding structure is lost. To counteract this loss, we introduce a method for pruning the document embeddings in their original high-dimensional space. Each core paper in the dataset is considered specialized within its field. Therefore, the embeddings of the new papers added through the citation network are evaluated for their proximity to each of the core paper embeddings. The intuition is that the embeddings of relevant papers should reside "close" to one or more of the embeddings of the core papers. Each core embedding is treated as the center of a hypersphere with radius $\rho$. 

\begin{figure}[ht]
    \centering
    \includegraphics[width=.8\linewidth]{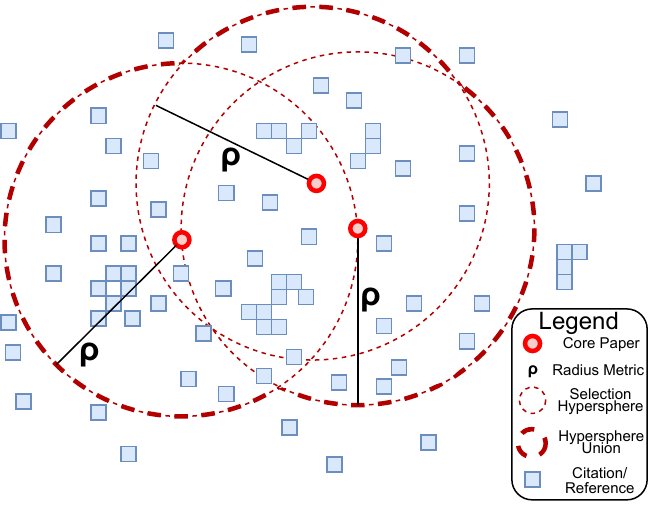}
    \caption{Hypersphere pruning calculation. Radii are compared from core paper distances, and once $\rho$ is selected, papers beyond the perimeter are pruned.}
    \label{fig:3}
\end{figure}

In mathematical notation, given a set of core papers $C = \{c_1, c_2, ..., c_n\}$ and their corresponding embeddings $E = \{e_1, e_2, ..., e_n\}$, the radius $r$ for each hypersphere is calculated as : \textit{First}, compute the pairwise Euclidean distances for all embeddings in $E$, forming a set $D = {d(e_i, e_j) : e_i, e_j \in E, i \neq j}$ where $d(e_i, e_j)$ denotes the Euclidean distance between embeddings $e_i$ and $e_j$. \textit{Second}, the median Euclidean distance between all core embeddings, denoted as $\rho$, where $\rho = \text{median}(D)$, can then be used as a threshold for including newly cited documents. The embedding for each core paper becomes a center of a hypersphere with radius $\rho$. A document embedding within one or more hyperspheres is at least as close to one or more core document embeddings as the median separation of the core document embeddings. Figure~\ref{fig:3} illustrates the process simplified to a 2-dimensional space.  



\subsection{Pruning through Topic Modeling} To further ensure topic cohesion, we perform topic modeling on the pruned dataset formed in the previous two steps. We utilize Semantic non-negative matrix factorization with automatic model selection (SeNMFk)\cite{eren2022senmfk}
for topic modeling. Given the documents dataset, we form a  term frequency-inverse document frequency (TF-IDF) 
matrix $\mat{X} \in \rm I\!R_{+}^{m \times n}$ and an SPPMI 
matrix $\mat{S} \in \rm I\!R_{+}^{m \times m}$ which encodes the semantic structure of the data (where $m$ is the number of tokens in the vocabulary and $n$ is the number of documents). We then jointly factorize $\mat{X}$ and $\mat{S}$ to produce two non-negative factor matrices $\mat{W} \in \rm I\!R_{+}^{m \times k}$ and $\mat{H} \in \rm I\!R_{+}^{k \times n}$, such that $\mat{X}_{ij} \approx \sum^{k}_{s} \mat{W}_{is} \mat{H}_{sj}$. Here, $\mat{W}$ represents the distribution of words across different topics, and $\mat{H}$ describes how these topics are distributed across the documents. After applying NMF, the information in $\mat{H}$ is used to associate each document with the topic it contributes most, forming clusters of documents. Only the documents corresponding to the topic, which comprises the core documents, are preserved.

It is important to conduct robust pre-processing of the documents to limit the noise introduced by expanding the dataset to produce a meaningful vocabulary. Our pre-processing procedure removes common stop-words, symbols, newline characters, HTML tags, non-ASCII characters, e-mail addresses, and copyright statements. Documents in languages other than English are identified and removed using heuristics such as the ratio of non-ASCII to total characters and the occurrence of common English stop-words in the text. There are instances where specific tokens or phrases denote unique terms in the chosen domain. While these terms might appear in different forms (such as spelling, acronym, or hyphenation), all forms signify the same concept. Standard preprocessing may split a multi-token term into separate tokens, which can destroy potentially crucial meaning. However, given  that an SME initially chooses the core papers, the SME can also pinpoint important terms and their assorted forms. Once these terms are identified, we consolidate all forms of each term into a singular entity. In the case of multi-token terms, we retain either the acronym or a hyphenated version  to ensure that the term's meaning is preserved in the TF-IDF and SPPMI matrices. In our tensors literature example, we substitute \textit{tensor-train} with \textit{\{TT, tensor train\}} and \textit{partial-differential-equation} with \textit{PDE} and all other various forms. Another strategy we employ at this pruning step to reduce noise involves reusing the same vocabulary for every hop. The vocabulary, derived from the core papers is consistently applied at each pruning decomposition. Consequently, less relevant papers (those using a significantly different vocabulary than the core) are represented as sparse entries in the TF-IDF matrix, reducing their influence on the decomposition. This step also enhances computational efficiency as the vocabulary dimension remains constant and does not grow with the number of documents.

Through these methods, BUNIE effectively enhances the thematic coherence of the dataset while maintaining topical alignment with the original core. This results in a significantly larger, interconnected dataset that retains the integrity of the original subject matter, ready for more in-depth exploration or application. Furthermore, to quantify the efficacy of our approach, we employed a compactness score, which is a metric that evaluates how closely the documents in the dataset are related to each other in terms of the topics they cover. The compactness score of a dataset is calculated using cosine similarity between the document embeddings. In mathematical terms, given a set of document embeddings $E = \{e_1, e_2, ..., e_n\}$, the compactness score $C$ 
is given by:

\begin{equation}
C = \frac{1}{n(n-1)} \sum_{i=1}^{n} \sum_{j=1, j\neq i}^{n} \frac{e_i \cdot e_j}{{|e_i|}_2 {|e_j|}_2}
\end{equation}

where $n$ is the total number of documents, $e_i$ and $e_j$ are the embeddings of the $i^{th}$ and $j^{th}$ document, $\cdot$ denotes the dot product, and $|\cdot|_2$ denotes the Euclidean norm. In measuring topic coherence using document embeddings, the cosine similarity between two embeddings, which ranges between -1 and 1, provides a measure of semantic alignment. A negative cosine similarity score, implying that the documents are semantically opposed, is an unlikely scenario within a specific topic. Therefore, we constrain the compactness score to fall between 0 and 1 to facilitate a meaningful quantification of topic coherence or alignment, accomplished by taking the absolute value of the cosine similarity. Higher values suggest a greater topic similarity between documents.
The final compactness score, a value also ranging between 0 and 1, is computed as the average cosine similarity across all pairs of documents in the dataset. By this measure, a higher compactness score indicates a more coherent or well-aligned set of documents regarding their topical content.

%% file: sec_results.tex
This section presents two experimental uses of BUNIE.
\begin{figure*}[ht!]
\centering
\begin{subfigure}{.45\textwidth}
    \centering
    \includegraphics[width=.95\linewidth]{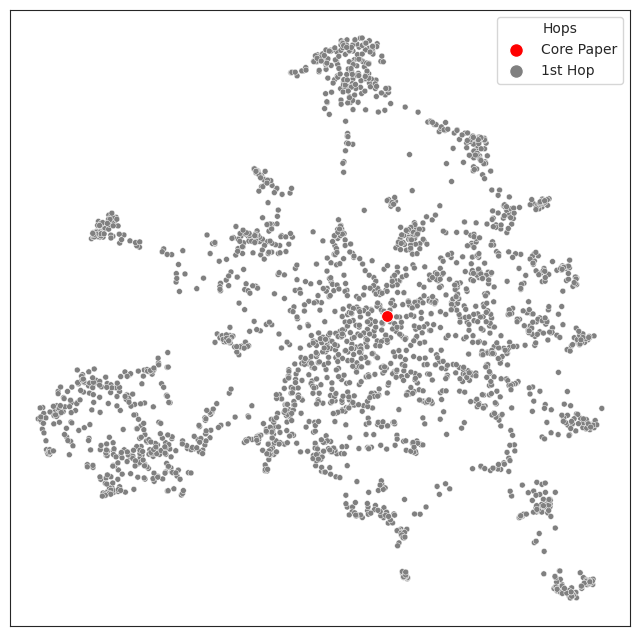}  
    \caption{}
    \label{SUBFIGURE LABEL 1}
\end{subfigure}
\begin{subfigure}{.45\textwidth}
    \centering
    \includegraphics[width=.95\linewidth]{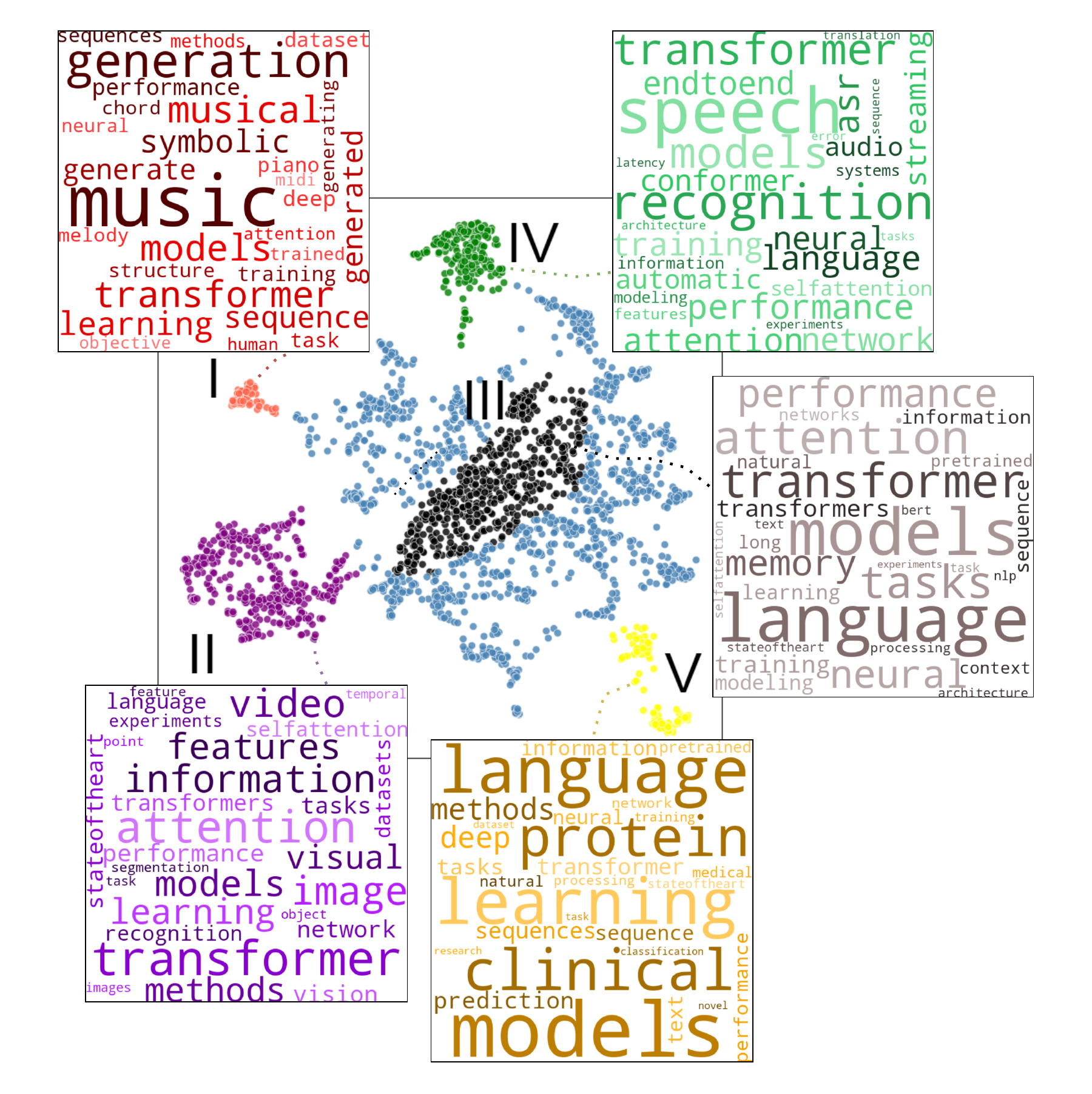}  
    \caption{}
    \label{SUBFIGURE LABEL 2}
\end{subfigure}
\caption{ Exploration of reference and citation paper topics out of an influential transformer paper \\ (a) ``Transformer-XL: Attentive Language Models Beyond a Fixed-Length Context'' in red,  citations/references in grey
\\ (b)  Manually selected paper clusters \& wordclouds from \ref{SUBFIGURE LABEL 1} colored: Music, Video, Transformer Performance, Speech, Proteins
}
\label{FIGURE LABEL}
\end{figure*}
\subsection{Expanding Targeted Dataset}


We first applied BUNIE to 10 papers hand-picked by an SME on a specific topic. These publications were influential papers in solving integral equations using tensor-train decomposition. With the "core" established, we sought to expand the dataset along the citation network. After the first hop, 632 citing papers were found. Using the visualization tool, we could quickly locate and prune papers that did not match the topic. While these papers tangentially addressed tensor decomposition, they failed to engage with the specific issues highlighted in the core papers. We then applied the automatic pruning through embeddings and SeNMFk pruning. After pruning the first hop, we were left with 411 papers, including the original 10 core papers. 

For such a minimal subset of papers, it was feasible to use the two-dimensional projection of the document embeddings in conjunction with bag-of-words word clouds to promptly identify the outlying papers. However, upon the second citation network expansion, the dataset grew rapidly to more than 8,000 papers. At this stage, the automatic pruning of the citation network became paramount. After pruning the second hops papers, a third hop was performed. After pruning, the final result came to 3,915 papers. This data flow demonstrates how BUNIE effectively combines human intuition with algorithmic utility to create a focused, relevant scientific dataset. 

As demonstrated in Table~\ref{table:tensors_compactness}, BUNIE's iterative process of topic expansion and alignment increases the compactness score of the dataset. While the first expansion to the citation network added many new documents to the dataset, it also introduced many unrelated documents, causing the compactness score to drop from 0.894 to 0.823.
The subsequent automatic pruning based on hypersphere proximity to the core document embeddings was effective in increasing the compactness score to 0.860, by eliminating less relevant documents, reducing the total document count to 4625. Following the hypersphere pruning, we carried out topic alignment by applying SeNMFk decomposition and selecting relevant subtopics, further refining the dataset. This increased the compactness score and resulted in a more manageable dataset containing 3915 documents. The increase in compactness score at each stage of the BUNIE process demonstrates the method's effectiveness in maintaining topic cohesion while expanding the dataset from a small set of core papers. 


 \begin{table}[ht]
 \caption{Compactness Score - Tensors}
 \centering
 \begin{tabular}{lcc} 
 \toprule
 Dataset & Compactness & Num. Documents \\ 
 \midrule
 Core Papers & 0.894 & 10 \\
 3-Hops, No Pruning & 0.823 & 10338  \\
 3-Hops, After Hypersphere Pruning  & 0.860 & 4625 \\
 3-Hops, After SeNMFk Pruning & 0.861 & 3915 \\
 \bottomrule
\end{tabular}
 \label{table:tensors_compactness}

\end{table}

\begin{table}[ht]
 \caption{Compactness Score - Audio processing}
 \centering
 \begin{tabular}{lcc} 
 \toprule
 Dataset & Compactness & Num. Documents \\ 
 \midrule
 Core Papers & 0.913 & 64 \\
 4-Hops, No Pruning & 0.798 & 15294 \\
 4-Hops, After Hypersphere Pruning & 0.861 & 1987 \\
 4-Hops, After SeNMFk Pruning & 0.861 & 1081 \\
 \bottomrule
\end{tabular}
 \label{table:music_compactness}

\end{table}
\subsection{Exploratory Data Expansion}

In recent years, the paper ``Transformer-XL: Attentive Language Models Beyond a Fixed-Length Context'' \cite{transformerXL} has drawn significant attention and influence across multiple research domains. Given this impact, it becomes interesting to explore the different domains influenced by the paper either individually or in relation to each other. BUNIE allows us to perform this exploration through topic modeling and visualizing text embedding projections. 

As demonstrated in \ref{SUBFIGURE LABEL 2},  we identified five prominent clusters associated with the following topics: audio processing, computer vision, speech processing, natural language processing, and proteins. From the visualized clusters, cluster I, containing music terms from 68 papers, was expanded through four hops along the citation and reference network, resulting in a significantly larger dataset comprising 15,294 papers. 
Following the expansion, the dataset was pruned through hyper-sphere calculation, retaining only papers within at least one of the 64 first-hop paper hyperspheres. At this point, the dataset contained 1,987 papers. Next, SeNMFk decomposed the papers into their core topic clusters, preserving 1,081 papers through 19 clusters, where only eight contained core papers and were preserved as the final dataset. The top words from the retained clusters in order were: music, attention, generative, lyric, video, score, learn, and emotion. As Table \ref{table:music_compactness} shows, the compactness of the dataset increased with each pruning step. The compactness of the core papers was 0.913, which decreased to 0.798 after the four-hop expansion due to the introduction of less-relevant papers. After hypersphere pruning, the compactness increased to 0.861, indicating the successful removal of off-topic papers. Remarkably, the compactness remained stable after SeNMFk pruning, suggesting that the most relevant papers were retained.

Notably, retained paper distributions per hypersphere pruned embedding mappings and SeNMFk decompositions will not always align with a human curator's intuitive UMAP-reduced selections. 
The discrepancy highlights the unique value of human judgment with algorithmic tools in dataset curations.

%% file: sec_conclusion.tex
This work contributes a novel system to build scientific datasets. With minimal input, we are able to iteratively build a dataset of scientic literature anchored on the core subject provided by an SME. At each step, the dataset is enlarged through the citation network and subsequently pruned using three separate methods, including one with human-in-the-loop. The result is an expanded dataset of work relevant to the core.

Promising future work is to seed an initial topic specification. The system would then iterate autonomously, filtering out documents and recalculating topic estimates to achieve topic distillation based on reinforcement learning. Auto-distillation could dynamically adapt the topic extraction and refinement based on continuous feedback on the topic's state. The system's efficiency and accuracy could improve over time, leading to more precise and reliable topic distillation. 

Additional considerations for future work include methods of forming the embeddings and creating a 'synthetic' core paper to serve as a foundation for automated topic alignment. The utilization of graph neural networks for understanding the relationship between the citations can also be explored, offering further insights into the structure and interconnections of the scientific literature. These enhancements and additions would augment the effectiveness and flexibility of BUNIE, further assisting researchers in their quest for knowledge.